\documentclass[aps,pra,preprint,floatfix,longbibliography, superscriptaddress]{revtex4-1}

\usepackage{pifont}
\usepackage[dvipsnames]{xcolor}
\definecolor{bananayellow}{rgb}{1.0, 0.88, 0.21}
\definecolor{amethyst}{rgb}{0.6, 0.4, 0.8}
\definecolor{ao(english)}{rgb}{0.0, 0.5, 0.0}

\usepackage{float}
\usepackage{epsfig}
\usepackage{bm}
\usepackage[matrix,frame,arrow]{xy}
\usepackage[applemac]{inputenc}
\usepackage[T1]{fontenc}
\usepackage{lmodern}
\usepackage[english]{babel}
\usepackage{amsmath}
\usepackage{ae}
\usepackage{amssymb}
\usepackage{color}
\usepackage{graphicx}
\usepackage{bbm}
\usepackage{url}
\usepackage{nomencl}
\usepackage{subfigure}
\usepackage{slashed}

\usepackage{fixmath}
\usepackage{booktabs}

\renewcommand{\eqref}[1]{\mbox{Eq.~(\ref{#1})}}

\newcommand{\be}{\begin{equation}}
\newcommand{\ee}{\end{equation}}
\newcommand{\bea}{\begin{eqnarray}}
\newcommand{\eea}{\end{eqnarray}}

\usepackage{xr}
\usepackage[colorlinks]{hyperref}
\hypersetup{%
    plainpages=true,
    breaklinks=true,
    hypertexnames=false,
    pageanchor=true,
    colorlinks=true,
    linkcolor={blue},
    citecolor={red},
    urlcolor={blue},
    anchorcolor={black}
}

\newcommand{\beq}{\begin{eqnarray}}
\newcommand{\eeq}{\end{eqnarray}}

\begin{document}
	
	
	\title{Non-perturbative Dynamical Casimir Effect \\in Optomechanical Systems: \\ Vacuum Casimir-Rabi Splittings}
	
	
	\author{Vincenzo Macr\`{i}}
	\affiliation{Dipartimento di Scienze Matematiche e Informatiche, Scienze Fisiche e  Scienze della Terra,
		Universit\`{a} di Messina, I-98166 Messina, Italy}
	\affiliation{Center for Emergent Matter Science, RIKEN, Saitama
		351-0198, Japan}
	\author{Alessandro Ridolfo}
	\affiliation{Center for Emergent Matter Science, RIKEN, Saitama
		351-0198, Japan}
	\author{Omar Di Stefano}
	\affiliation{Dipartimento di Scienze Matematiche e Informatiche, Scienze Fisiche e  Scienze della Terra,
		Universit\`{a} di Messina, I-98166 Messina, Italy}
	\affiliation{Center for Emergent Matter Science, RIKEN, Saitama
		351-0198, Japan}
	\author{Anton Frisk Kockum}
	\affiliation{Center for Emergent Matter Science, RIKEN, Saitama 351-0198, Japan}
	
	\author{Franco Nori}
	\affiliation{Center for Emergent Matter Science, RIKEN, Saitama
		351-0198, Japan} \affiliation{Physics Department, The University
		of Michigan, Ann Arbor, Michigan 48109-1040, USA}
	
		\author{Salvatore Savasta}
	\affiliation{Dipartimento di Scienze Matematiche e Informatiche, Scienze Fisiche e  Scienze della Terra,
		Universit\`{a} di Messina, I-98166 Messina, Italy}
	\affiliation{Center for Emergent Matter Science, RIKEN, Saitama
		351-0198, Japan}


	\begin{abstract}
		
		We study the dynamical Casimir effect using a fully quantum-mechanical description of both the cavity field and the oscillating mirror. We do not linearize the dynamics, nor do we adopt any parametric or perturbative approximation. 
		By numerically diagonalizing the full optomechanical Hamiltonian, we show that the resonant generation of photons from the vacuum is determined by a ladder of mirror-field {\em vacuum Rabi splittings}. We find that vacuum emission can originate from the free evolution of an initial pure mechanical excited state, in analogy with the spontaneous emission from excited atoms.
		By considering a coherent drive of the mirror, using a master-equation approach to take losses into account, we are able to study the dynamical Casimir effect for optomechanical coupling strengths ranging from weak to ultrastrong.
		We find that a resonant production of photons out of the vacuum can be observed even for mechanical frequencies lower than the cavity-mode frequency. Since high mechanical frequencies, which are hard to achieve experimentally, were thought to be imperative for realizing the dynamical Casimir effect, this result removes one of the major obstacles for the observation of this long-sought effect. We also find that the dynamical Casimir effect can create entanglement between the oscillating mirror and the radiation produced by its motion in the vacuum field, and that vacuum Casimir-Rabi oscillations can occur.
		
	\end{abstract}
	
	
	\maketitle
	
	
	Quantum-field theory predicts that vacuum fluctuations can be converted into real particles by the energy provided through certain external perturbations \cite{Schwinger1951,Moore1970,Hawking1974,Schwinger1993,Yablonovitch1989,Wilson2011,Laehteenmaeki2013,Nation2012}.
	Examples include the Schwinger effect \cite{Schwinger1951}, predicting the production of electron-positron pairs from the vacuum under the application of intense electrical fields; Hawking radiation \cite{Hawking1974,Hawking1975}, which is caused by the bending of space-time in intense gravitational fields and determines the evaporation of black holes; the Unruh effect \cite{Unruh1976}, predicting that an accelerating observer will observe blackbody radiation where an inertial observer would observe none; and the dynamical Casimir effect (DCE) \cite{Moore1970,Wilson2011,Laehteenmaeki2013}  describing the generation of photons from the quantum vacuum due to rapid changes of the geometry (in particular, the positions of some boundaries) or material properties of electrically neutral macroscopic or mesoscopic objects.
	
	The creation of photons by moving mirrors was first
    predicted by Moore \cite{Moore1970} in 1970, for a one-dimensional cavity. In 1976, Fulling and Davis \cite{Fulling1976} demonstrated that photons can be generated even by a single mirror, when it is subjected to a nonuniform acceleration.
	Since the first prediction of the DCE, many different experimental setups, able to produce sudden nonadiabatic changes inducing light emission from the quantum vacuum, have been proposed \cite{Dodonov2010}. These proposals can be divided  into two main groups: setups where the photons are created due to the movement of mirrors \cite{Fulling1976,Sassaroli1994,Lambrecht1996,Dodonov1996,Schaller2002,Kim2006} [mechanical (M) DCE], and systems where the boundary conditions are modulated by some effective motion producing a parametric amplification of vacuum fluctuations \cite{Wilson2011,Laehteenmaeki2013,Nation2012,Uhlmann2004, Crocce2004, Braggio2005, Segev2007, DeLiberato2007, Johansson2009, Johansson2010} [parametric (P) DCE].
	 
	 The experimental detection of the DCE is challenging owing  to the difficulty in changing the boundary conditions, e.g., by moving physical objects, such as massive mirrors, sufficiently fast for generation of a significant number of photons. 
	 In 1996, Lambrecht, Jaekel, and Reynaud \cite{Lambrecht1996} provided a quantitative estimate of the photon flux radiated from an optomechanical system consisting of a cavity with oscillating mirrors. Taking advantage of  resonance-enhancement effects, they showed that a significant number of microwave photons, sufficient to allow detection, can be produced in realistic  high-$Q$ cavities with a moderate peak velocity of the mirrors. However, the  resonance condition, requiring that the mechanical oscillation frequency $\omega_{\rm m}$ be at least twice that of the lowest-frequency cavity mode $\omega_{\rm c}$, remains a major barrier to the experimental demonstration of the MDCE. Recently, high-frequency mechanical oscillators, [$\omega_{\rm m} / (2 \pi) \sim 6$ GHz] \cite{OConnell2010} have been realized. However, to produce vacuum radiation at a frequency $\omega_{\rm c} / (2 \pi)\sim 5$ GHz, a still higher mechanical frequency $\omega_{\rm m} / (2 \pi) \sim 10$ GHz  is required.
	 
	 In order to circumvent
	 these difficulties, a number of theoretical proposals has suggested
	 to use experimental setups where the boundary conditions are
	 modulated by some effective motion instead. Examples of such
	 proposals include: (i) using lasers to rapidly modulate the reflectivity of thin semiconductor films \cite{Crocce2004, Braggio2005}, (ii) modulating the resonance frequency of a superconducting stripline  resonator \cite{Segev2007}, non-adiabatic time-modulation  of (iii) the light-matter coupling strength  in cavity QED systems \cite{DeLiberato2007,DeLiberato2009,Garziano2013, Garziano2014,Hagenmuller2016, Cirio2016a,Delib2016}, or (iv) of the background in which the field propagates \cite{Yablonovitch1989,Uhlmann2004,Laehteenmaeki2013}, and (v) to use a superconducting quantum interference device (SQUID)
	 to modulate the boundary condition of a superconducting
	 waveguide or resonator \cite{Johansson2009,Johansson2010,Wilson2011}.
	 
	 Recently, using superconducting circuits \cite{you2005,You2011,xiang2013},  the DCE (specifically, the PDCE) has been demonstrated experimentally, implementing proposals (iv) and (v). In particular,  it was observed \cite{Wilson2011} in a coplanar transmission line terminated by a SQUID whose inductance was modulated at high frequency ($> 10$ GHz). This was also demonstrated by modulating the index of refraction of a Josephson metamaterial embedded in a microwave cavity \cite{Laehteenmaeki2013}.
	 
	 Most theoretical studies of the MDCE  consider the mirror that scatters the vacuum field to follow a prescribed motion \cite{Moore1970,Fulling1976, Sassaroli1994, Dodonov1996,
	 	Plunien2000, Haro2006, Dodonov1998, Dodonov2010}.
	 Therefore the photon creation from the initial vacuum state is usually described as a parametric amplification process, just as in the case of the PDCE. 
	 Exceptions consider fluctuations of the position of the mirror driven by vacuum radiation pressure using linear dispersion theory \cite{Jaekel1993,Maia1993} or focus on the mirror motion as the main dynamical degree of freedom. In the latter case, studies have shown how the DCE induces friction forces on the mirror \cite{Kardar1999,Maghrebi2013} or leads to decoherence of mechanical quantum superposition states \cite{Dalvit2000}.
	 	 	\begin{figure}
	 	 	\centering
	 	 	\includegraphics[width = 14 cm]{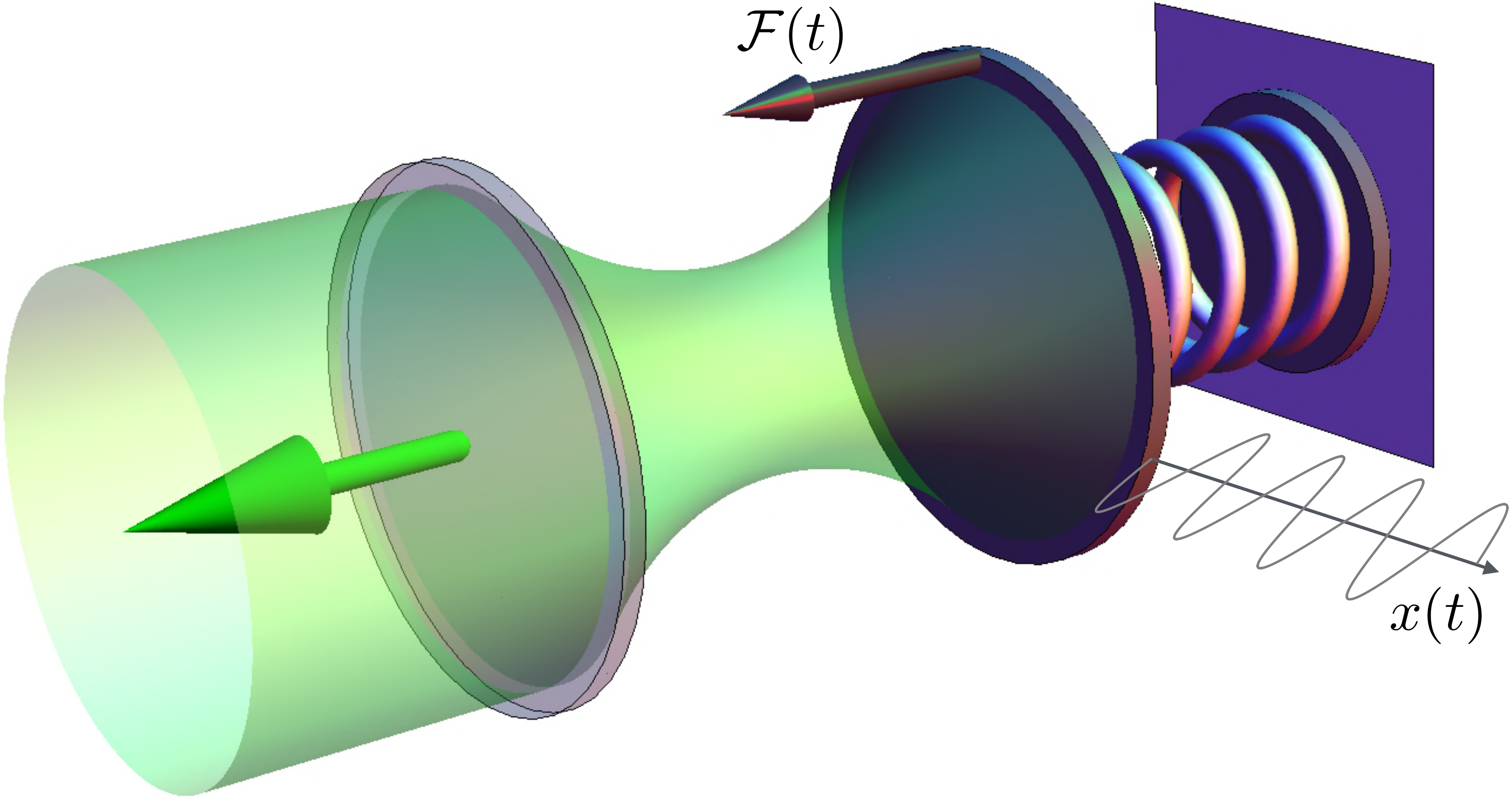}
	 	 	\caption{Schematic of a generic optomechanical system where one of the mirrors in an optical cavity can vibrate at frequency $\omega_{\rm m}$. If the vibrating mirror is  excited by an external drive ${\cal F}(t)$, able to create a $k$-phonon state with a non-negligible probability, the vibrating mirror can emit photon pairs if $k \omega_{\rm m} \simeq 2 \omega_{\rm c}$, where $\omega_{\rm c}$ is the resonance frequency of the cavity.
	 	 		\label{fig:1}}
	 	 \end{figure}
	 
	 In this article, we investigate the MDCE in cavity optomechanical systems \cite{Aspelmeyer2014}, 
	 treating both the cavity field and the moving mirror as quantum-mechanical systems. Calculations are made {\it without}  performing any linearization of the dynamical equations. 
	 Multiple scattering between the two subsystems is taken into account nonperturbatively. The interactions of the two subsystems with the environment is described by using a master equation  \cite{Breuer2002,Hu2015}. A surprising feature of this approach is that the DCE can be described  {\it without} the need for a time-dependent light-matter interaction. The only time-dependent Hamiltonian term considered here is  the one describing the external drive of the vibrating mirror. Actually, within this approach, the DCE effect can be described, at least in principle, even {\it without} considering any time-dependent Hamiltonian. Specifically,  we find that vacuum radiation can originate from the free evolution of  an initial pure mechanical excited state, in analogy with the spontaneous emission from excited atoms. We believe that this theoretical framework provides a more fundamental explanation of the DCE. Note that fundamental processes in quantum field theory are described by interaction Hamiltonians which do {\it not} depend parametrically on time.
	 
  We find that the resonant generation of photons from the vacuum is determined by a ladder of mirror-field vacuum Rabi-like energy splittings. When the loss rates are lower than the corresponding frequency splittings, a reversible exchange of energy between the vibrating mirror and the cavity field, which we call {\em vacuum Casimir-Rabi oscillations}, can be observed.

	   Cavity-optomechanics experiments are rapidly approaching the regime where the radiation pressure of a single photon displaces the mechanical oscillator by more than its zero-point uncertainty \cite{Teufel2011,Chan2011,Rimberg2014,Heikkila2014,Pirkkalainen2015,Nation2016,Rouxinol2016}. Specifically, in circuit optomechanics, it has been shown that the radiation pressure effect can be strongly enhanced by  introducing a qubit-mediated \cite{Heikkila2014,Pirkkalainen2015} or  modulated \cite{Liao2014} interaction  between the mechanical and the electromagnetic resonator. 
	   This ultrastrong coupling (USC) regime, where the optomechanical coupling rate is comparable to the mechanical frequency, can give rise to strong nonlinearities
	  even in systems described by the standard optomechanics interaction Hamiltonian, which depends linearly on the mirror displacement \cite{Nunnenkamp2011,Garziano2015b}. 
	  This regime favours the  observation of macroscopic mechanical oscillators in a nonclassical state of motion \cite{Nunnenkamp2011, Stannigel2012,Garziano2015b, Macri2016}. This requires a full quantum treatment of both the mechanical and optical degrees of freedom, and  multiple-scattering effects between the field and the mechanical oscillator cannot be ignored. 
	  
	  The approach considered here allows us to extend the investigation of the DCE to the USC limit of cavity optomechanics. We find that this regime is able to remove one of the major obstacles for the experimental observation of this long-sought effect. Indeed, we show that, approaching USC, a resonant production of photons out from the vacuum can be observed for mechanical frequencies lower than the lowest cavity-mode frequency. Approximately, the resonance condition for the production of photon pairs out from the vacuum is $k\, \omega_{\rm m} \simeq  2 \omega_{\rm c}$, with $k$ integer. This corresponds to processes where $k$ phonons in the mechanical oscillator are converted into two cavity photons. The matrix element for this transition decreases rapidly for increasing $k$, but increases when the optomechanical coupling $g$ increases. Already the resonance condition with $k=2$, corresponding to $\omega_{\rm m} \simeq  \omega_{\rm c}$, where DCE matrix elements display reasonable amplitude even at moderate coupling,  is promising for the observation of the MDCE. 
	   Indeed, this resonance condition can  be achieved in the GHz spectral range using ultra-high-frequency mechanical micro- or nano-resonators \cite{OConnell2010, Rouxinol2016}. 
	  
	  Very recently, new resonance conditions in the DCE that potentially allow the production of photons for $\omega_{\rm m} < \omega_{\rm c}$ have been found assuming a classical  prescribed anharmonic motion of the mirror \cite{Ordaz-Mendoza2016}. This model, however, describes photon emission from vacuum fluctuations only in the instability region and the resulting time evolution of the mean photon number grows exponentially even in the presence of cavity losses. On the contrary, in our approach, the vibrating mirror is treated as a harmonic (anharmonicity only originates from the interaction) quantum degree of freedom on the same footing as
	   the cavity field and we do not find unstable regions.
	  
	
	\section{Results}
	\subsection{Model}
	\label{model}

	We consider the case of a cavity with a movable end mirror (see Fig.~\ref{fig:1}) and focus on the simplest possible model system in cavity optomechanics, which has been used to successfully describe most of such experiments to date.
	A detailed derivation of the optomechanical Hamiltonian can be found in Ref.~\cite{Law1995}.
	Both the cavity field and the position of the mirror are treated as dynamical
	variables and a canonical quantization procedure is adopted.
	By considering only one mechanical mode with resonance frequency $\omega_{\rm m}$ and bosonic operators $\hat b$ and $\hat b^\dag$, and only the lowest-frequency optical mode $\omega_{\rm c}$ of the cavity, with bosonic operators $\hat a$ and $\hat a^\dag$, the system Hamiltonian can be written as 
	$\hat H_{\rm s} = \hat H_0 + \hat H_{\rm I}$, 
	where
	\begin{equation}\label{H0}
	\hat H_0 = \hbar \omega_{\rm c} \hat a^\dag \hat a +  \hbar \omega_{\rm m} \hat b^\dag \hat b
	\end{equation}
	is the unperturbed Hamiltonian. The Hamiltonian describing the mirror-field interaction is
	\begin{equation}
	\hat H_{\rm I} = \frac{\hbar G}{2} (\hat a + \hat a^\dag)^2 \hat x\, ,
	\end{equation}
	where $\hat x = x_{\rm ZPF}(\hat b + \hat b^\dag)$ is the  mechanical displacement ($x_{\rm ZPF}$ is the zero-point fluctuation amplitude of the vibrating mirror) and $G$ is a coupling parameter. By developing the photonic factor in normal order, and by defining 
	new bosonic phonon and photon operators and a renormalized photon frequency, $\hat H_{\rm s}$ can be written 
	as  
	\begin{equation}\label{Hs}
	\hat H_{\rm s} = \hat H_0 + \hat V_{\rm om} + \hat V_{\rm DCE}\, ,
	\end{equation}
	where $\hat H_0$ formally coincides with Eq.~(\ref{H0}),
	\begin{equation}
	\hat V_{\rm om} =\hbar g \hat a^\dag \hat a(\hat b + \hat b^\dag)
	\end{equation}
	is the standard optomechanical interaction conserving the number of photons, and
	\begin{equation}\label{VDCE}
	\hat V_{\rm DCE} = \frac{\hbar g}{2} (\hat a^2 + \hat a^{\dag 2} ) (\hat b + \hat b^\dag)\, ,
	\end{equation}
	describes the creation and annihilation of photon pairs \cite{Butera2013}, where $g = G\, x_{\rm ZPF}$ is the optomechanical coupling rate.
	As we will see in detail below, $\hat V_{\rm DCE}$ determines the DCE. The Hamiltonian (\ref{Hs}) describes the interaction between a moving mirror and the radiation pressure of a cavity field. However, the same radiation-pressure-type coupling is obtained for microwave optomechanical circuits (see, e.g., Ref.~\cite{Heikkila2014}).

	When describing most of the optomechanics experiments to date \cite{Aspelmeyer2014}, $\hat V_{\rm DCE}$ is  neglected. This is a very good approximation when the mechanical frequency is much smaller than the cavity frequency (which is the most common experimental situation), since  $\hat V_{\rm DCE}$ connects bare states with an energy difference $2 \hbar \omega_{\rm c} \pm \hbar \omega_{\rm m}$ much larger than the coupling strength $\hbar g$. With this approximation, the resulting Hamiltonian, $\hat H_0 + \hat V_{\rm om}$, conserves the number of photons and can be analytically diagonalized. 
	The full Hamiltonian in Eq.~(\ref{Hs}) provides the simplest unified description of cavity-optomechanics experiments and the DCE in a cavity with a vibrating mirror.
	
	In order to properly describe the system dynamics, including external driving, dissipation and decoherence, the coupling to external degrees of freedom needs to be considered.
	A coherent external drive of the vibrating mirror can be described by including the following time-dependent Hamiltonian,
	\be\label{F}
	\hat V_{\rm m}(t) = {\cal F}(t)\, (\hat b + \hat b^\dag)\, ,
	\ee
	where ${\cal F}(t)$ is proportional to the external force applied to the mirror.
	Analogously, the coherent optical excitation of the cavity mode can be described by
	\be
	\hat V_{\rm c}(t) = {\cal E}(t)\,  (\hat a + \hat a^\dag)\, ,
	\ee
	where ${\cal E}(t)$ is proportional to the coherent optical field exciting the cavity. In the following, we will only consider the external excitation of the mirror [${\cal E}(t) = 0$] by a continuous-wave drive or by a pulse, in contrast to most cavity optomechanical experiments, where the system is optically excited.

	\subsection{Vacuum Casimir-Rabi splittings }
	\label{Rabi}
	\begin{figure*}
		\centering
		\includegraphics[width = \linewidth]{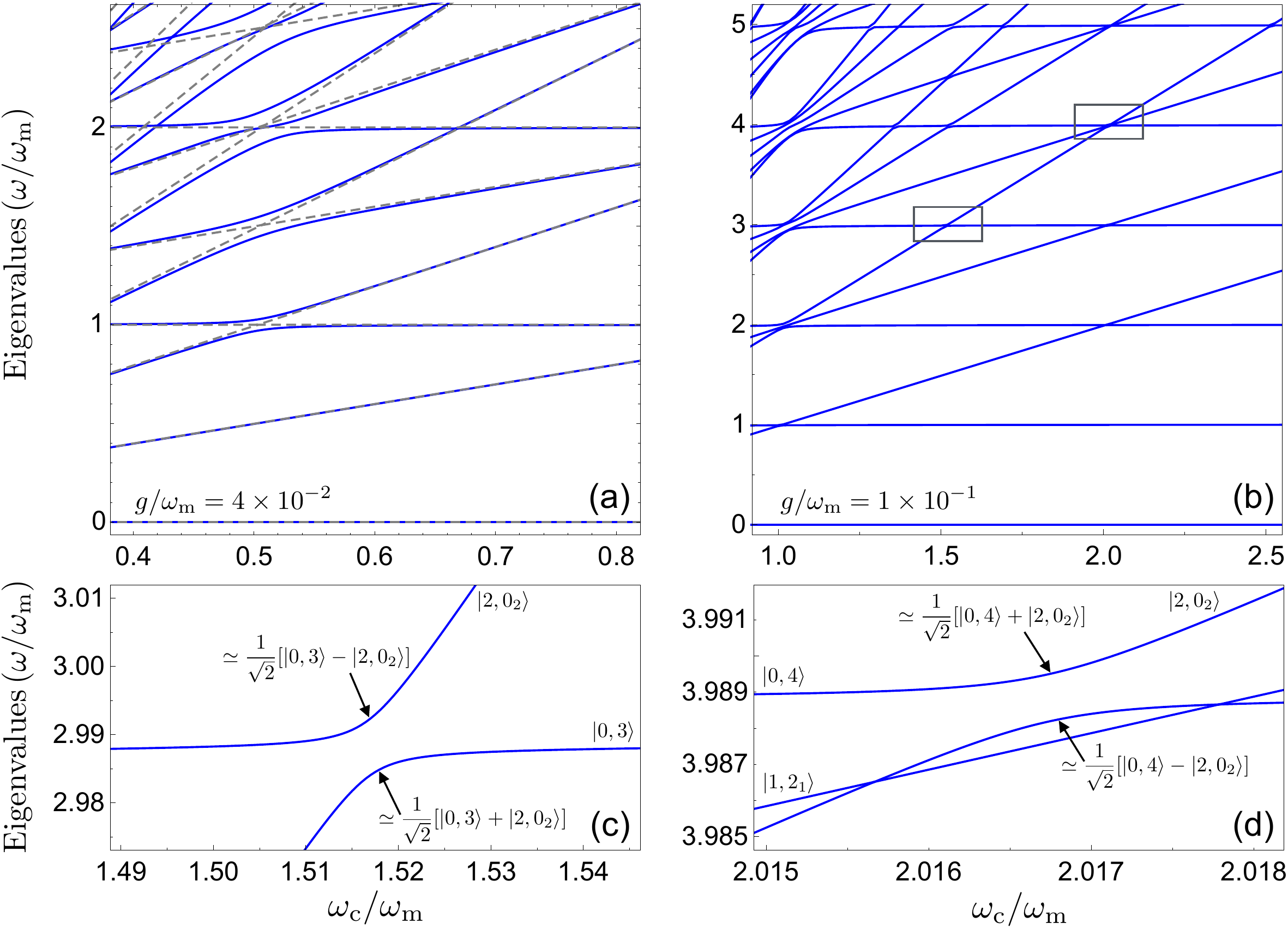}
		\caption{Lowest energy levels of the system Hamiltonian as a function of the ratio between the cavity frequency and the mechanical frequency. In (a)  an optomechanical coupling $g/\omega_{\rm m} = 0.04$ has been used. The dashed grey lines describe the eigenenergies of the standard optomechanics Hamiltonian $\hat H_{0} + \hat V_{\rm om}$. The blue continuous curves are the eigenvalues of $\hat H_{\rm s} = H_{0} + \hat V_{\rm om} + \hat V_{\rm DCE}$, which  have also been calculated for a coupling $g/\omega_{\rm m} = 0.1$ as shown in (b). Panels (c) and (d) display enlarged views of the two boxed regions in (b) showing avoided level crossings due to optomechanical hybridizations of zero- and two-photon states (vacuum Casimir-Rabi splittings).
			\label{fig:2}}
	\end{figure*}

We begin by numerically diagonalizing the system Hamiltonian $\hat H_{\rm s}$ in Eq.~(\ref{Hs}). Figure~\ref{fig:2}(a) displays the lowest energy levels as a function of the ratio between the cavity and the mechanical frequency, using an optomechanical coupling $g/\omega_{\rm m} = 0.04$.
For comparison, we also show in Fig.~\ref{fig:2}(a) (dashed grey lines) the lowest energy levels $$E_{n,k} = \hbar \omega_{\rm c} n - \hbar g^2 n^2 / \omega_{\rm m} + \hbar \omega_{m} k$$ for the standard optomechanics Hamiltonian $\hat H_{0} + \hat V_{\rm om}$.
In this case, the system eigenstates can be written as 
\begin{equation}
	|n, k_{n}\rangle = |n \rangle_{\rm c} \otimes  \hat D(n \beta)|k \rangle_{\rm m}\, ,
\end{equation}	
where $n$ is the cavity photon number and the mechanical state $|k_n \rangle$ is a displaced Fock state, determined by the displacement operator $\hat D(n \beta)= \exp[{n \beta (\hat b^\dag - \hat b)}]$, with $\beta = g/\omega_{\rm m}$. 
The dashed grey horizontal lines in Fig.~\ref{fig:2}(a) correspond to  states $|0, k_{0}\rangle \equiv |0, k\rangle$ belonging to the $n=0$ manifold. The dashed grey lines with lower non-zero slope (slope $1$) describes the $n=1$ manifold ($|1, k_{1}\rangle$), while those with slope $2$ describe the energy levels of the manifold with $n=2$.

The continuous blue lines correspond to the energy levels obtained by  numerically diagonalizing the system Hamiltonian $\hat H_{\rm s}$ in Eq.~(\ref{Hs}). The main difference compared to the grey lines is the appearance of level anticrossings of increasing size at increasing eigenenergy values when $E_{0,k} = E_{2, k-1}$ (corresponding to a cavity frequency  $\omega_{\rm c} = \omega_{\rm m}/2 + 2 g^2/\omega_{\rm m} \simeq \omega_{\rm m}/2$). We observe that the condition
$\omega_{\rm c} \simeq \omega_{\rm m}/2$ is the standard
resonance condition ($\omega_{\rm m} = 2 N \omega_{\rm c}$)
for the DCE in a cavity with a vibrating mirror \cite{Lambrecht1996}, with $N=1$.
These avoided crossings arise from the coherent coupling induced by $\hat V_{\rm DCE}$ between the states $|0,k \rangle \leftrightarrow |2,(k-1)_2 \rangle$ with $k \geq 1$.
If the optomechanical coupling is not too strong, the size of the anticrossings can be analytically calculated by using first-order perturbation theory.
By approximating $|k_2\rangle \simeq  |k \rangle$, for the energy splittings, we obtain the simple expression $$2 \hbar \Omega_{0,k}^{2,k-1} = 2  \langle 2, (k-1)_2| \hat V_{\rm DCE} | 0, k \rangle \simeq \hbar g \sqrt{2k},$$ in very good agreement with the numerical results in Fig.~\ref{fig:2}(a). When the splitting is at its minimum ($\omega_{\rm c} = \omega_{\rm m}/2 + 2 g^2/\omega_{\rm m}$), the two system eigenstates are approximately (not exactly, owing to dressing effects induced by $\hat V_{\rm DCE}$) the symmetric and antisymmetric superposition states
\be
	| \psi_{2(3)} \rangle \simeq \frac{1}{\sqrt{2}} (|0,1 \rangle \pm |2, 0_{2} \rangle )\, .
\ee
These {\em vacuum Casimir-Rabi splittings}, demonstrating optomechanical-induced hybridization of zero- and two- photon states,  establish a close analogy between the DCE and cavity QED, where the atom-photon vacuum Rabi splitting and  quantum Rabi oscillations in the time domain have been observed in many systems and widely exploited for many applications \cite{Haroche2013}. We observe, however, that, while quantum Rabi splittings in cavity QED describe coherent coupling between states with the same number of excitations, Casimir-Rabi splittings involve pairs of states with  different number of excitations. In this case, for example, a state with $k$ excitations (phonons) hybridizes with a state with $k+1$ excitations  ($k-1$ phonons and $2$ photons). This non-conservation of excitation numbers is reminiscent of cavity QED in the USC regime, where the counter-rotating terms in the atom-field interaction Hamiltonian enable the coupling of states with different excitation numbers \cite{Niemczyk2010, Ma2015, Garziano2015, Garziano2016, Kockum2017, Stassi2017}. More generally, the DCE Hamiltonian in Eq.~(\ref{VDCE}) gives rise to several additional avoided-level crossings, describing resonant optomechanical scattering processes $| n, k_n \rangle \leftrightarrow |n+2, (k- q)_{n+2} \rangle$, which  occur when the energies of the final and initial states coincide ($2 \omega_{\rm c} \sim q\,  \omega_{\rm m}$).

These coherent couplings induced by $\hat V_{\rm DCE}$ constitute the fundamental quantum mechanism through which mechanical energy is transferred to the vacuum electromagnetic field, giving rise to the DCE. For example, in the absence of losses, an initial 1-phonon-0-photon state $|0,1 \rangle$ (not being a system  eigenstate)
will evolve as
\be\label{23}
| \psi(t) \rangle = \cos{(\Omega_{0,1}^{2,0}\, t)}\, |0,1 \rangle - i  \sin{(\Omega_{0,1}^{2,0}\, t)}\, |2,0_2 \rangle\, ,
\ee 
and thus, after a time $t= \pi/(2 \Omega_{0,1}^{2,0})$, will spontaneously evolve into a photon pair.
This elementary analysis shows that, if the mechanical and photonic loss rates are much lower than the coupling rate 
$\Omega_{0,1}^{2,0}$, mechanical energy can be converted, at least in principle, into light with $100 \%$ efficiency. Moreover, according to Eq.~(\ref{23}), at $t= \pi/(4 \Omega_{0,1}^{2,0})$, the moving mirror and the cavity field become maximally entangled. 

Figure~\ref{fig:2}(b) shows the lowest energy eigenvalues of $\hat H_{\rm s}$ for larger cavity-mode frequencies, using a stronger optomechanical coupling $g/\omega_{\rm m} = 0.1$.
The figure [see also the boxed details enlarged in panels~\ref{fig:2}(c) and (d)] shows that 
optomechanical resonant couplings occur also for $\omega_{\rm m} < 2 \omega_{\rm c}$. In particular,  vacuum Casimir-Rabi splittings occur when
$E_{0,k} = E_{2, k-q}$, corresponding to a cavity frequency 
$\omega_{\rm c} - 2 g^2/\omega_{\rm m} \simeq  q\, \omega_{\rm m}/2$, also when $q > 1$.
Avoided level crossings for $q=2$ are clearly visible in Fig.~\ref{fig:2}(b). Smaller splittings for $q= 3$ and $q = 4$ are indicated by black boxes and their enlarged views are shown in Figs.~\ref{fig:2}(c) and (d).
By using first-order perturbation theory, the size of these avoided level crossings can be calculated analytically:
\be\label{V}
2 \hbar \Omega_{0,k}^{2,k-q} = 2  \langle 0, k| \hat V_{\rm DCE} | 2, (k-q)_2 \rangle 
= \sqrt{2}\,  \hbar g\left[\sqrt{k+1}  D_{k+1, k-q}(2 \beta) 
+  \sqrt{k} D_{k-1, k-q}(2 \beta) \right]\, , 
\ee
where the matrix elements of the displacement operators can be expressed in terms of associated Laguerre polynomials: $D_{k',k}(\alpha)= \sqrt{k!/k'!} \alpha^{k'-k} e^{-|\alpha|^2/2} L_k^{k'-k}(|\alpha|^2)$\, . We note that the resonance conditions with $\omega_{\rm m} \leq \omega_{\rm c}$ have nonzero DCE matrix elements  (\ref{V}) thanks to the non-orthogonality of mechanical Fock states with different phonon numbers, belonging to different photonic manifolds: $$\langle k | k_2^\prime \rangle = D_{k, k^\prime}(2 \beta) \neq 0$$ (see, e.g., Ref.~\cite{Macri2016}). Note also that, for $\beta \to 0$, $\langle k | k_2^\prime \rangle \to \delta_{k,k^\prime}$. Examples of analytically-calculated splittings $2  \Omega_{0,k}^{2,k-q}$  are displayed in the Methods section, where we also present a comparison between the numerically-calculated  vacuum Rabi splitting  and the corresponding analytical calculations, obtained with first-order perturbation theory  for $2 \Omega_{0,3}^{2,0}$ and $2 \Omega_{0,4}^{2,0}$. 

Also for resonance conditions with $q >1$, when the splitting is at its minimum (corresponding to values of $\omega_{\rm c}$ such that $E_{0,k} \simeq E_{2, k-q}$), the two system eigenstates are essentially symmetric and antisymmetric linear superpositions.
For example, for the boxed splitting at lower $\omega_{\rm c}$,
\be\label{5-6}
| \psi_{5(6)} \rangle \simeq \frac{1}{2} (|0,3 \rangle \pm |2, 0_{2} \rangle )\, .
\ee
Neglecting losses, an initial 3-phonon state $|0,3 \rangle$ (not being a system  eigenstate) will thus evolve spontaneously as
\be\label{23t}
| \psi(t) \rangle = \cos{(\Omega_{0,3}^{2,0}\, t)}\, |0,3 \rangle - i  \sin{(\Omega_{0,3}^{2,0}\, t)}\, |2,0 \rangle\, ,
\ee
giving rise to a $100 \%$  mechanical-to-optical energy transfer and to vacuum-induced entanglement.

\subsection{DCE in the weak-coupling regime}
	\label{weak}
	
Here we investigate the  dynamics giving rise to the DCE, numerically solving the system master equation (described in Methods).  We focus on some experimentally promising cases, with  $\omega_{\rm c} \geq \omega_{\rm m}$.  
\begin{figure}
	\centering
	\includegraphics[width = 14 cm]{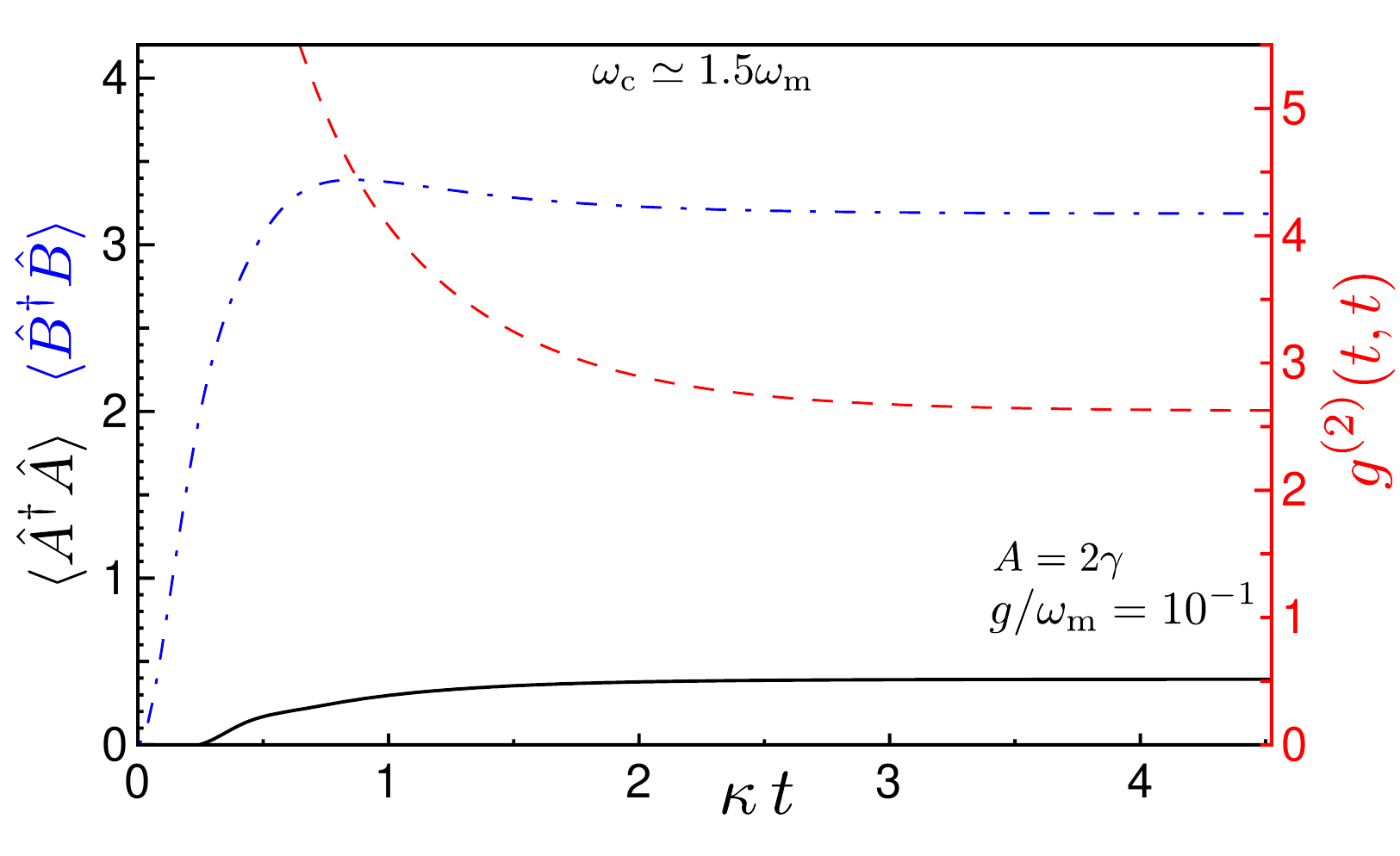}
	\caption{System dynamics for $\omega_{\rm c} \simeq 1.5 \omega_{\rm m}$ under continuous-wave drive of the vibrating mirror. The blue dash-dotted curve describes the mean phonon number $\langle \hat B^\dag \hat B \rangle$, while the black solid curve describes the mean intracavity photon number $\langle \hat A^\dag \hat A \rangle$ rising thanks to the DCE. The zero-delay normalized photon-photon correlation function $g^{(2)}(t,t)$ is also plotted as a red dashed curve with values given on the $y$-axis on the right. All parameters are given in the text.
		\label{fig:3}}
\end{figure}
\begin{figure}
	\centering
	\includegraphics[width = 14 cm]{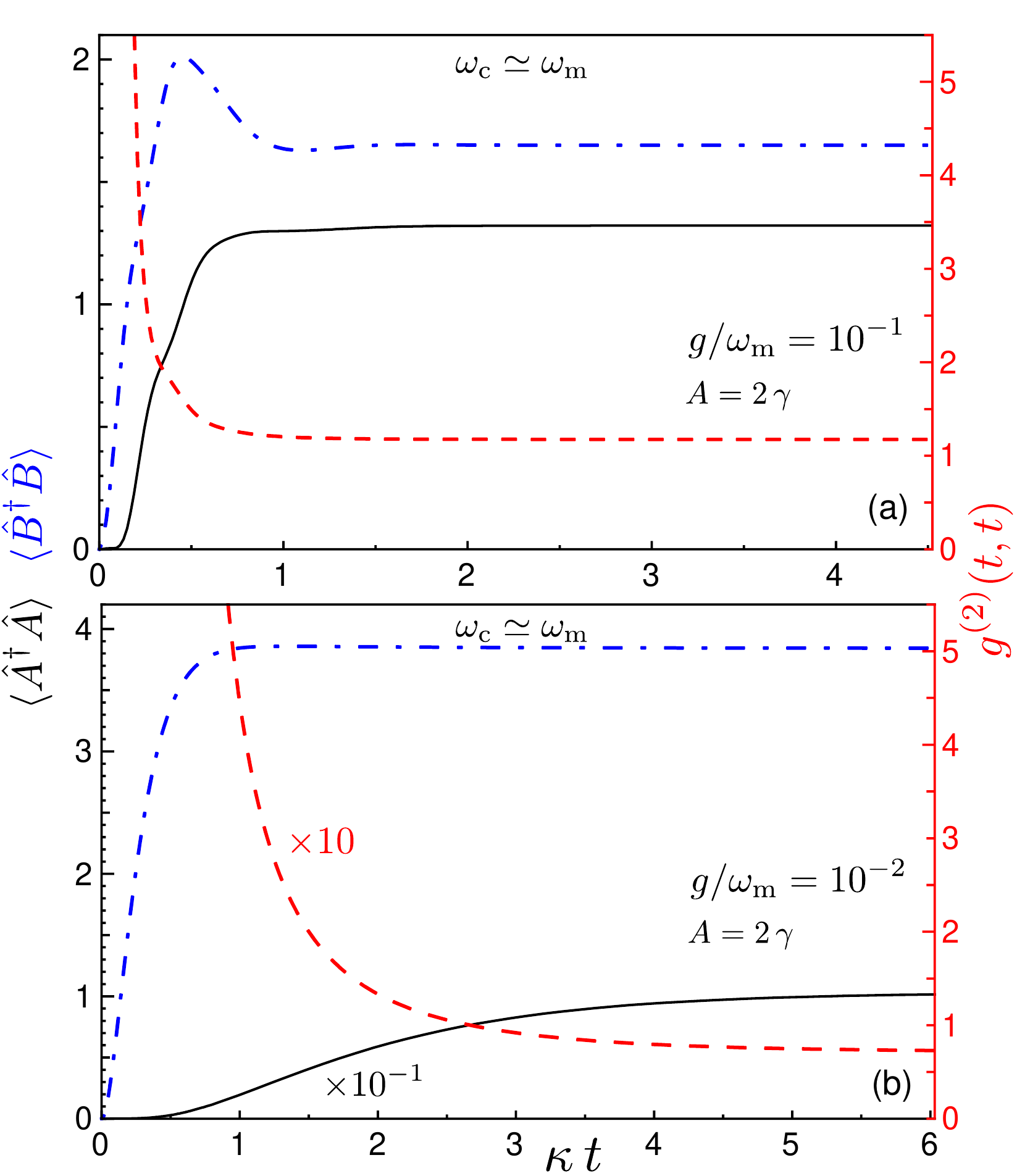}
	\caption{System dynamics for $\omega_{\rm c} \simeq  \omega_{\rm m}$ under continuous-wave drive of the vibrating mirror. The blue dash-dotted curves describe the mean phonon number $\langle \hat B^\dag \hat B \rangle$, while the black solid curves describe the mean intracavity photon number $\langle \hat A^\dag \hat A \rangle$ rising thanks to the DCE. The zero-delay normalized photon-photon correlation function $g^{(2)}(t,t)$ is also plotted (red dashed curve with values given on the $y$-axis on the right). Panel (a) has been obtained using $g /\omega_{\rm m} = 0.1$; panel (b) with $g /\omega_{\rm m} = 10^{-2}$. All the other parameters are given in the text.
		\label{fig:4}}
\end{figure}
In this subsection, we limit our investigations to the {\em weak-coupling regime}, which, however, does not refer to the optomechanical coupling strength (we use $g/ \omega_{\rm m}$ up to $0.1$).  Instead,  following the terminology of cavity QED, by the term {\em weak}, we mean  Casimir-Rabi splittings $2\Omega_{0,k}^{2,k-q}$ smaller than the total decoherence rate $\gamma + \kappa$, where $\gamma$ and $\kappa$ are the mechanical and photonic loss rates respectively (see Methods).
We consider the optomechanical system initially in its ground state and numerically solve the master equation (\ref{ME}) including the excitation of the moving mirror by a single-tone continuous-wave mechanical drive ${\cal F}(t) = A \cos{(\omega_{\rm d} t)}$. Figure~\ref{fig:3} shows the time evolution of the mean phonon number $\langle \hat B^\dag \hat B \rangle$ (blue dash-dotted curve), the  intracavity mean photon number $\langle \hat A^\dag \hat A \rangle$ (black solid curve), and the equal-time photonic normalized second-order correlation function (red dashed curve)
\be
g^{(2)}(t,t) = \frac{\langle \hat A^\dag(t) \hat A^\dag(t) \hat A(t) \hat A(t) \rangle}{\langle \hat A^\dag(t) \hat A(t) \rangle^2 }\, ,
\ee  
where $\hat A, \hat B$ are dressed photonic and phononic operators, as explained in the Methods section.
We assume a zero-temperature reservoir and use  $\kappa/ \omega_{\rm m} = 3 \times 10^{-3}$ and $\gamma = 10 \kappa$ for the photonic and mechanical loss rates. We consider a weak ($A/\gamma = 2$) resonant excitation of the vibrating mirror ($\omega_{\rm d} = \omega_{\rm m}$). 
Figure~\ref{fig:3} shows the system dynamics for the case $\omega_{\rm c} \simeq 3 \omega_{\rm m}/2$ [corresponding to the minimum level-splitting shown in  Fig.~\ref{fig:2}(c)]. We used a normalized coupling $g/ \omega_{\rm m} = 0.1$. The results demonstrate that a measurable rate of photons is produced. In particular, a steady-state mean intra-cavity photon number $\langle \hat A^\dag \hat A \rangle_{\rm ss} \simeq 0.3$ is obtained, corresponding (for a resonance frequency of the cavity mode $\omega_{\rm c}/ (2\pi) = 6$ GHz) to a steady-state output photon flux $\Phi = \kappa \langle \hat A^\dag \hat A \rangle_{\rm ss} \sim 3 \times 10^6$ photons per second. The output photon flux is remarkable, taking into account the weak mechanical drive, corresponding to a steady-state mean phonon number  $\langle \hat b^\dag \hat b \rangle_{\rm ss} = 4$ for $g/{\omega_{\rm m}}= 0$, and the quite low cavity quality factor $Q_{\rm c} =  \omega_{\rm c} / \kappa = 500$ used in the numerical calculations. Note that $Q_{\rm c}$-values beyond $10^6$ are obtained in microwave resonators (see, e.g., \cite{Megrant2012}). Also, the mechanical loss rate $\gamma$ used here corresponds to a quality factor $Q_{\rm m}$ one order of magnitude lower than the  experimentally measured values in ultra-high-frequency mechanical resonators \cite{OConnell2010, Rouxinol2016}. Such a low driving amplitude and quality factors were used in order to reduce both memory and numerical effort.
We observe that the steady-state phonon number does not reach the value $\langle \hat B^\dag \hat B \rangle_{\rm ss} = 4$ obtained in the absence 
of $\hat V_{\rm DCE}$. This is an expected result, since the calculations fully take into account the correlated field-mirror dynamics induced by the DCE.
The calculated second-order correlation function  $g^{(2)}(t,t)$, also displayed in Fig.~\ref{fig:3}, starts with very high values, confirming that photons are emitted in pairs. As time goes on, it decreases significantly, due to losses which affect the photon-photon correlation and to the increase in the mean photon number (note that $g^{(2)}(t,t)$, owing to the squared denominator, is intensity dependent). 

Figure~\ref{fig:4} displays results for the case $\omega_{\rm c} \simeq \omega_{\rm m}$. In this case, a higher-frequency mechanical oscillator is required. However, as we pointed out in the introduction, mechanical oscillators with resonance frequencies  $\omega_{\rm m} / (2 \pi) \sim 6$ GHz  have been realized \cite{OConnell2010}. In the present case, the DCE can be observed by coupling such a mechanical oscillator to a microwave resonator with the same resonance frequency. The advantage of this configuration is that the corresponding matrix elements (vacuum Casimir-Rabi splittings) $\Omega^{2, k-2}_{0,k}$ are non-negligible even for quite low optomechanical couplings. Figure~\ref{fig:4}(a), obtained using a coupling $g /\omega_{\rm m} =0.1$, shows a remarkable energy transfer from the moving mirror to the cavity field which in its steady state contains more than 1 photon, corresponding to a steady-state output photon flux $\Phi$ beyond $10^7$ photons per second. Figure~\ref{fig:4}(b) has been obtained using an optomechanical coupling one order of magnitude lower. The resulting steady-state 
 mean intra-cavity photon number decreases by one order of magnitude to $\langle \hat A^\dag \hat A \rangle_{\rm ss} \simeq 0.1$, but it is still measureable. This result is particularly interesting because it shows that the MDCE can be observed with state-of-the-art ultra-high-frequency mechanical resonators \cite{OConnell2010, Rouxinol2016} and with normalized coupling rates $\beta$ below those already achieved in circuit optomechanics \cite{Teufel2011}.
We can conclude that the MDCE can be observed at $\omega_{\rm c} \simeq \omega_{\rm m}$ even when the optomechanical USC regime is not reached, although reaching it can significantly enhance the emission rate. 

	\begin{figure*}
	\centering
	\includegraphics[width = \linewidth]{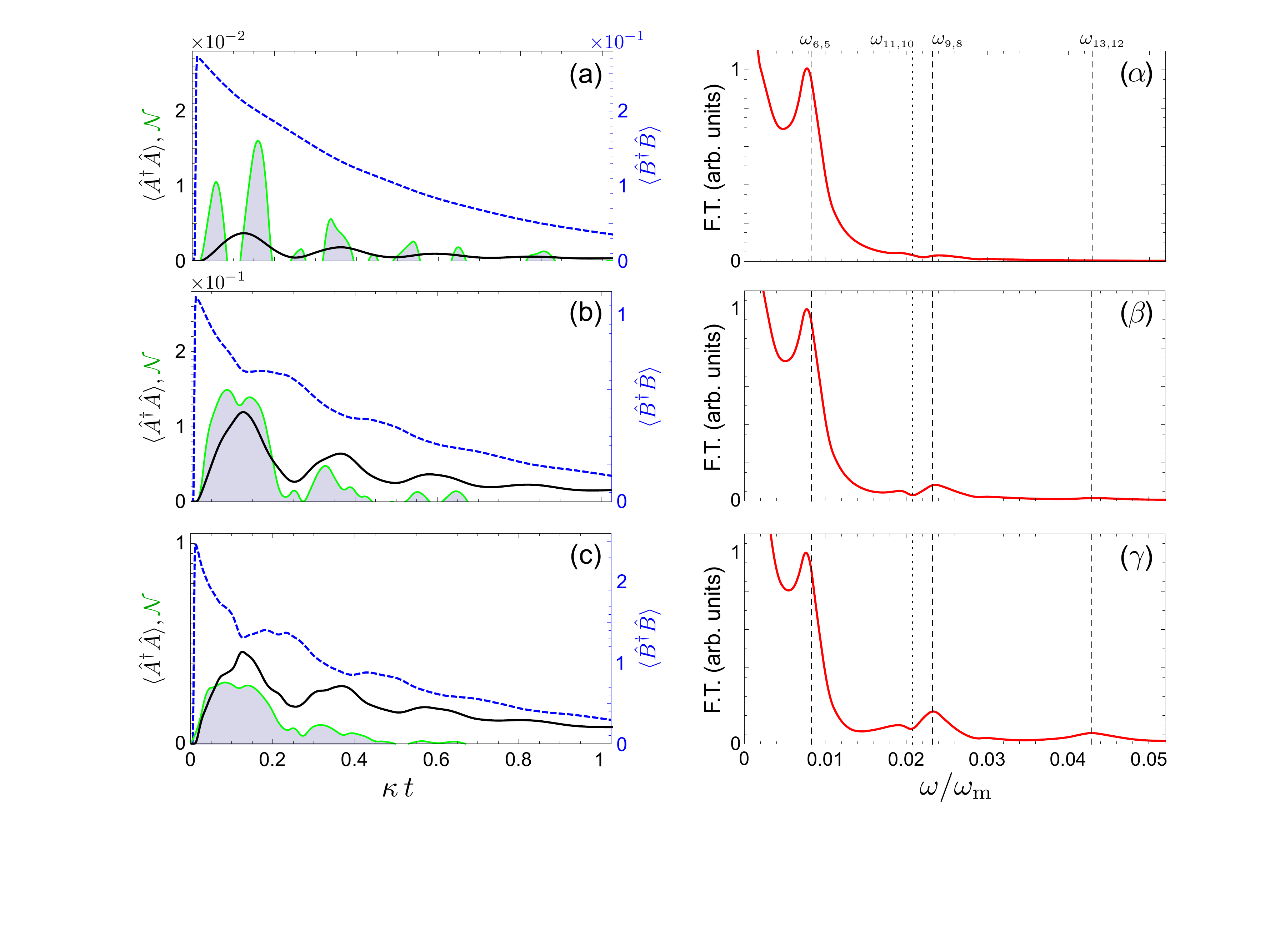}
	\caption{System dynamics after the pulse arrival, obtained for pulses with amplitudes increasing from top to bottom: ${\cal A} = \pi /3$ (a), $2 \pi /3$ (b), and $\pi$ (c). Specifically, panels (a-c) display the mean intracavity photon number $\langle \hat A^\dag \hat A \rangle$ (black solid curves), the mean phonon number $\langle \hat B^\dag \hat B \rangle$ (blue dashed curves), and the negativity ${\cal N}$ (green filled curve). Panels ($\alpha$-$\gamma$) display the Fourier transform of the mean photon number shown in the corresponding panel on the left.
		\label{fig:5}}
\end{figure*}

Quantum correlations in microwave radiation produced by the DCE in
a superconducting waveguide terminated and modulated by a SQUID have been investigated \cite{Johansson2013}. The results indicate that the produced radiation can be strictly
nonclassical and can display a measurable amount of intermode entanglement.
In the approaches where the real or effective mirror is assumed to follow a prescribed classical motion, the entanglement between the moving mirror and the emitted electromagnetic field cannot be investigated. On the contrary, the present theoretical framework, fully taking into account the quantum correlations between the moving mirror and the cavity field, induced by $\hat V_{\rm DCE}$, allows us to investigate if the DCE creates optomechanical  entanglement.
In the present case, the dynamics involve many system states and, owing to the presence of losses, the system is far from being in a pure state during its time evolution. We quantify the entanglement using the negativity ${\cal N}$ (see Methods).
By considering the same, numerically calculated, density matrix  used to derive the results shown in Fig.~\ref{fig:3}, we find a steady-state negativity oscillating around ${\cal N} \simeq 5 \times 10^{-2}$, attesting that the DCE is able to produce mirror-field steady-state entanglement.
For comparison, a maximally entangled Bell-like state, like that described by Eq.~(\ref{23t}) at time $t = \pi / (4 \Omega_{0,3}^{2,0})$, has a negativity ${\cal N} =0.5$.
 Using the parameters of Fig.~\ref{fig:4}(a), we find a larger steady-state negativity, oscillating around ${\cal N} \simeq 0.1$.
However, we find no entanglement  (${\cal N} \simeq 0$) for the parameters of Fig.~\ref{fig:4}(b), when the influence of the DCE on the dynamics of the moving mirror is small.

\subsection{Vacuum Casimir-Rabi oscillations}
\label{osc}

Here we investigate the DCE in the {\em strong-coupling regime}, when the  Casimir-Rabi splittings $2\Omega_{0,k}^{2,k-q}$ are larger than the total decoherence rate $\gamma + \kappa$. This regime is particularly interesting, since it provides direct evidence of the level structure determining the DCE and the multiple-scattering effects between the two sub-systems. Moreover, as we are going to show, it gives rise to non-perturbative entangled dynamics of the cavity field and the moving mirror.

We numerically solve the master equation (\ref{ME}), assuming the optomechanical system prepared in its ground state and including the vibrating mirror excitation by an ultrafast resonant pulse
${\cal F}(t) = {\cal A}\, {\cal G}(t-t_0) \cos{(\omega_{\rm d}\, t)}$,
where ${\cal G}(t)$ is a normalized Gaussian function. 
We consider an optomechanical coupling $g/ \omega_{\rm m} = 0.1$ and set the cavity frequency at the value providing the minimum level-splitting $2 \Omega^{2,0}_{0,3} / \omega_{\rm m} \simeq 8 \times 10^{-3}$ shown in  Fig.~\ref{fig:2}(c) ($\omega_{\rm c} \simeq 3 \omega_{\rm m}/2$).
We consider pulses with central frequency resonant with the mechanical oscillator ($\omega_{\rm d} = \omega_{\rm m}$), and with standard deviation $\sigma = (20 \Omega^{2,0}_{0,3} )^{-1} \simeq 12 / \omega_{\rm m}$. For the loss rates, we use  $\gamma = 0.15 \; \Omega^{2,0}_{0,3} \simeq 6 \times 10^{-4}\, \omega_{\rm m}$ and $\kappa = \gamma /2$.

Figure~\ref{fig:5} displays the system dynamics after the pulse arrival and the Fourier transform of the mean photon number, obtained for pulses with amplitudes increasing from top to bottom: ${\cal A} = \pi /3,\, 2 \pi /3,\, \pi$. Figures~\ref{fig:5}(a)-(c) show  (Casimir-Rabi) nutations (superimposed on the exponential decay due to the presence of losses) of the signals
$\langle \hat A^\dag \hat A \rangle$ (black solid curves) and
$\langle \hat B^\dag \hat B \rangle$ (blue dashed curves). The mean phonon number  displays much less pronounced oscillations which are anticorrelated with the photonic oscillations. When the pulse amplitude is small ($A = \pi /3$), the time evolution of the mean photon number  is  sinusoidal-like with peak amplitudes decaying exponentially. Initially, the ultrafast kick produces a coherent mechanical state. The lowest Fock state in this mechanical coherent superposition, which is able to resonantly produce photon pairs, is $|0, 3 \rangle$ (see Fig.~\ref{fig:2}). This state is coherently coupled to the state $|2, 0_2  \rangle$ by $\hat V_{\rm DCE}$, giving rise to the avoided-crossing states $| \psi_5 \rangle$ and $| \psi_6 \rangle$ given in Eq.~(\ref{5-6}). These two levels display a frequency vacuum Casimir-Rabi splitting   $2 \Omega^{2,0}_{0,3}$, which, as also shown in Fig.~\ref{fig:5}($\alpha$), corresponds to the frequency of the observed Rabi-like oscillations. The small difference between the peak in Fig.~\ref{fig:5}($\alpha$) and $\omega_{6,5} \equiv (E_6- E_5)/\hbar = 2 \Omega^{2,0}_{0,3}$ is due to the presence of the nearby higher peak at $\omega=0$. For the amplitude $A = \pi /3$,
 the peak phonon number, reached just after the kick, is significantly below one, and thus the occupation probability for the state  $|3  \rangle_{\rm m}$ is very low. This explains the weakness of the photonic signal in Fig.~\ref{fig:5}(a) and the smallness of the oscillations superimposed on the exponential decay in the mechanical signal $\langle \hat B^\dag \hat B \rangle$. Indeed, the mechanical states $|1  \rangle_{\rm m}$ and $|2  \rangle_{\rm m}$ in the initial coherent  superposition have a much larger probability than $|3  \rangle_{\rm m}$ and evolve unaffected by the vacuum field. Higher energy mechanical states $| k \rangle_{\rm m}$ with $k >3$ can also produce photon pairs at a rate $\Omega^{2,k-3}_{0,k}$, but their occupation probability is negligible at such a low pulse amplitude. The non-monotonous dynamics of the signals indicates that the DCE effect is, at least partially, a reversible process: the emitted photon pairs can be reabsorbed by the moving mirror and then re-emitted, if the effective DCE rates are larger than the losses. However, if one of the photons in the pair is lost, the surviving one-photon state is no longer resonant with the vibrating mirror and undergoes a standard exponential decay. This effect gives rise to a decay of the oscillation amplitude faster than the signal decay. 
 
 When increasing the pulse amplitude [Figs.~\ref{fig:5}(b) and \ref{fig:5}(c)], the mean photon number grows significantly and no longer oscillates sinusoidally. In addition, the mechanical signal deviates significantly from  the exponential decay, owing to increased population of the mechanical states with phonon number $k \geq 3$ that are able to produce photon pairs.  

Figures~\ref{fig:5}($\alpha$) to \ref{fig:5}($\gamma$) show the Fourier transforms of
the photonic nutation signals. For $A = \pi /3$, besides the intense peak at $\omega = 0$ (describing the exponential decay induced by losses, always superimposed on the nutations), only an additional peak at $\omega \simeq \Omega^{2,0}_{0,3}$ is visible, in full agreement with the sinusoidal signal in Fig.~\ref{fig:5}(a).
Increasing the pulse amplitude, a second peak at higher frequency [Fig.~\ref{fig:5}($\beta$)], followed by a third at still higher frequency [Fig.~\ref{fig:5}($\gamma$)] appears. These two additional peaks in the Fourier transform clarify the origin of the non-sinusoidal behaviour of signals in Figs.~\ref{fig:5}(b) and \ref{fig:5}(c). They correspond to the higher-energy processes associated with the effective coupling strengths $\Omega^{2,1}_{0,4}$ and $\Omega^{2,2}_{0,5}$, both larger than $\Omega^{2,0}_{0,3}$. However, these two peak frequencies are slightly larger than the corresponding minimum half-splittings $\Omega^{2,1}_{0,4}$ and $\Omega^{2,2}_{0,5}$. 
This difference occurs because the ladder of vacuum Casimir-Rabi splittings, occurring at a given cavity frequency when $ 2 \omega_{\rm c} \simeq 3 \omega_{\rm m}$, is not perfectly vertical (see Fig.~\ref{fig:2}), owing to energy shifts induced by $\hat V_{\rm DCE}$. Hence, if $\omega_{\rm c}$, as in this case, is tuned to ensure that the minimum level splitting  $E_6 - E_5 = 2 \hbar \Omega^{2,0}_{0,3}$, the higher-energy split levels will not be at their minimum. The peaks clearly visible in Fig.~\ref{fig:5}($\gamma$) occur at frequencies $\omega = 0$, $\omega \simeq \omega_{6,5} = 2 \Omega^{2,0}_{0,3}$, $\omega \simeq \omega_{9,8} > 2 \Omega^{2,1}_{0,4}$, and $\omega \simeq \omega_{13,12} > 2 \Omega^{2,2}_{0,5}$.
A further structure with a dip is also visible in Figs~\ref{fig:5}($\beta$) and \ref{fig:5}($\gamma$) around $\omega = \omega_{11,10}$. This corresponds to the coherent coupling of the states $|1, 3_1 \rangle$ and $|3, 0 \rangle$, producing the eigenstates $|\psi_{10} \rangle$ and $|\psi_{11} \rangle$. These states are neither directly excited by the external mechanical pulse which generates zero-photon states, nor by $\hat V_{\rm DCE}$, which creates or destroy photon pairs. However, the cavity losses can give rise to the decay $| 2, 3_1 \rangle \to | 1, 3_1 \rangle$. Hence, also the states $|\psi_{10} \rangle$ and $|\psi_{11} \rangle$ can be indirectly involved in the signal dynamics.

Analogous quantum Rabi oscillations, giving rise to discrete Fourier components, have been experimentally observed for circular Rydberg atoms in a high-$Q$ cavity \cite{Brune1996}. In this system, however, the different level anticrossings are not affected by different energy shifts.

In cavity QED, the strong-coupling dynamics produces atom-field entanglement \cite{Raimond2001}. We investigate if this non-perturbative regime of the DCE is able to produce entanglement between the mobile-mirror and the cavity field, when the mirror is excited by a coherent pulse and in the presence of mechanical and optical dissipations.
The time evolution of the negativity is displayed in Figs.~\ref{fig:5}(a)-(c). As expected, ${\cal N}$ increases noticeably when  the pulse amplitude  increases, so that the mirror dynamics is significantly affected by the DCE. 
We observe that,  while decaying as a consequence of losses, the negativity displays a non-monotonous behaviour analogous to that observed in cavity-QED \cite{LoFranco2007}.

\subsection{Radiative decay of a mechanical excited state}
\label{MSE}
Spontaneous emission is the process in which a quantum emitter, such as a  natural or an artificial atom, or a molecule, decays from an excited state to a lower energy state and emits a photon. This cannot be described within the classical electromagnetic theory and is fundamentally a quantum process. Here
we present numerical calculations showing that a vibrating mirror prepared in an excited state (mechanical Fock state) can spontaneously emit photons like a quantum emitter. In this case, however, instead of a single photon, a photon pair is emitted.
Here, instead of considering the coherent excitation of the vibrating mirror as in usual descriptions of the DCE, we assume that it is initially prepared in a Fock state. We consider the case $\omega_{\rm c} \simeq \omega_{\rm m}$ and the system is initialised in the state $|0, 2 \rangle$, with $\omega_{\rm c}$ sufficiently detuned from the DCE resonance (minimum avoided-level crossing) at  $\omega^0_{\rm c} \simeq \omega_{\rm m}$, with $\delta \omega_{\rm c} \equiv \omega_{\rm c} - \omega^0_{\rm c} = 0.1 \omega^0_{\rm c}$, such that the effective resonant DCE coupling is negligible. This  $k = 2$ mechanical Fock state can be prepared, for example, if the vibrating mirror is strongly coupled to an addtional qubit \cite{OConnell2010}, using the same protocols realized in circuit QED \cite{Hofheinz2009}. After preparation, the cavity can be quickly tuned into resonance: $\omega_{\rm c} \to \omega^0_{\rm c}$. If the cavity resonator is an LC superconducting circuit, its resonance frequency can be tuned by using a SQUID.
In order to not affect the mechanical Fock state during this non-adiabatic process, the tuning time  must be shorter than $2 \pi / \Omega_{0,2}^{2,0}$.
	\begin{figure}
	\centering
	\includegraphics[width = 11.5 cm]{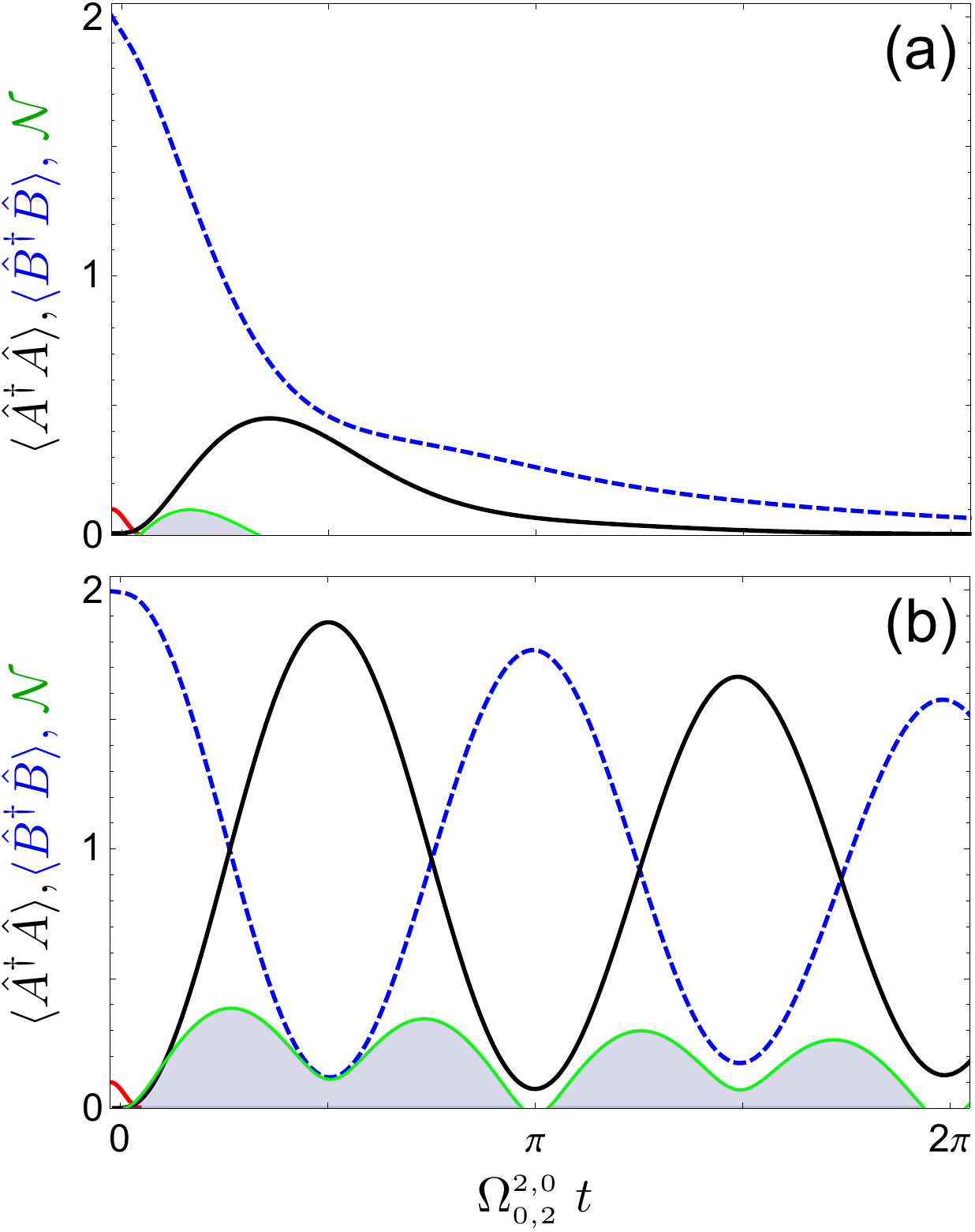}
	\caption{Dynamics starting from a mechanical Fock state. The blue dashed  curves decribe the mechanical signal $\langle \hat B^\dag \hat B \rangle$, while the black solid  curves describe the  optical signal $\langle \hat A^\dag \hat A \rangle$. The  green curves correspond to the negativity ${\cal N}$. The cavity frequency is initially detuned from the DCE resonance ($\delta \omega_{\rm c} = 0.1$), and the system is initially  prepared in the state $|0, 2 \rangle$. Then the cavity is quickly tuned to the DCE resonance ($\delta \omega_{\rm c} \to 0$). The initial detuning $\delta \omega_{\rm c}$  is displayed as a small red solid curve in the lower left corner of both panels, near $t=0$.
	In panel (a), the dynamics is evaluated in the weak-coupling regime. Panel (b) displays the vacuum Casimir-Rabi oscillations that arise when the system loss rates are low. The parameters used are provided in the text.
		\label{fig:6}}
\end{figure}

Figure~\ref{fig:6} displays the mean phonon number $\langle \hat B^\dag \hat B \rangle$ (dashed blue curve), the mean intracvity photon number  $\langle \hat A^\dag \hat A \rangle$ (black solid  curve), and the negativity (green filled curve) calculated for $g / \omega_{\rm m} = 0.1$. Figure~\ref{fig:6} also displays the initial detuning $\delta \omega_{\rm c}$ (red solid curve). Figure~\ref{fig:6}(a), obtained 
using $\gamma = \Omega_{0,2}^{2,0}/5$ and $\kappa = 2.5 \gamma$, describes the irreversible mechanical decay due to both non-radiative (induced by the mechanical loss rate $\gamma$) and radiative decay (induced by $\hat V_{\rm DCE}$). The radiative decay  gives rise to non-negligible light emission (black solid curve), and to transient mirror-field entanglement.
 Figure~\ref{fig:6}(b), obtained using the lower
 loss rates $\gamma = \kappa = \Omega_{0,2}^{2,0}/80$, shows vacuum Casimir-Rabi  oscillations. In this case, a photon pair can be produced at $t = \pi / (2 \Omega^{2,0}_{0,2})$ with probability close to one.

\section{Conclusions}
\label{C}
We have analyzed the DCE in cavity optomechanical systems, describing quantum-mechanically both the cavity field and the vibrating mirror, fully including multiple scattering between the two subsystems. The full quantum approach developed here describes the DCE {\it without} introducing a time-dependent light-matter interaction. The only time-dependent Hamiltonian term considered in this work was the one describing the external drive of the moving mirror. Actually, we can conclude that
 the DCE can be described even without considering any time-dependent Hamiltonian. Vacuum emission can originate from the free evolution of an initial pure mechanical excited state, in analogy with the spontaneous emission from excited atoms.
 
Using numerical diagonalization of the optomechanical Hamiltonian \cite{Law1995}, including those terms usually neglected for describing current optomechanics experiments \cite{Aspelmeyer2014}, we have shown that
the resonant generation of photons from the vacuum is determined by a ladder of mirror-field vacuum Rabi-like splittings. These avoided-level crossings describe the energy-conserving conversion of phonons (quanta of mechanical vibration) into photon pairs. More generally, the DCE Hamiltonian in Eq.~(\ref{VDCE}) describes many resonant optomechanical scattering processes $| n, k_n \rangle \leftrightarrow |n+2, (k- q)_{n+2} \rangle$ which occur when the energies of the final and initial states coincide ($2 \omega_{\rm c} \sim q\,  \omega_{\rm m}$).

The standard resonance condition for the DCE  requires a mechanical resonance frequency at least double that of the lowest mode frequency of the cavity. 
We have shown instead that, when the coupling between the moving mirror and the cavity field is non-negligible compared to the mechanical and optical resonance frequencies, a resonant production of photons out from the vacuum can be observed for mechanical frequencies equal to or lower than the cavity-mode frequencies.
Hence, the present analysis demonstrates that optomechanical systems with coupling strengths which experiments already started to approach, and with vibrating mirrors working in the GHz spectral range, can be used to observe light emission from mechanical motion.

We have also analyzed the non-perturbative regime of the DCE, which, we showed, provides direct access to the level structure determining the DCE effect and can display Rabi-like nutations of the cavity-field and oscillating mirror signals. Finally, we have shown that the oscillating mirror can evolve into a state which is entangled with the radiation emitted  by the mirror itself.

\section{Methods}
\label{methods}
\subsection{Master equation}

We take into account dissipation and decoherence effects by adopting a master-equation approach. For strongly-coupled hybrid quantum systems, the description offered by the standard quantum-optical master equation breaks down \cite{Beaudoin2011,Hu2015}. 
Following Refs.~\cite{Breuer2002,Ma2015, Hu2015}, we express the system-bath interaction Hamiltonian in the basis formed by the energy eigenstates of $\hat H_{\rm s}$.  By applying the standard Markov approximation and tracing out the reservoir degrees of freedom, we arrive at the master equation for the density-matrix operator $\hat \rho(t)$,
\begin{equation}\label{ME}
\dot{\hat\rho}(t) = \frac{i}{\hbar} [\hat \rho(t),\hat H_{\rm s}+ \hat V_{\rm m}(t)] +
\kappa\, {\cal D}[\hat A] \hat \rho(t) + \gamma\, {\cal D}[\hat B] \hat \rho(t)\, .
\end{equation}
Here the constants $\kappa$ and $\gamma$ correspond to the cavity-field and mirror damping rates. The dressed photon and phonon lowering operators  $\hat O = \hat A, \hat B$ are defined in terms of their
bare counterparts $\hat o = \hat a, \hat b$ as \cite{Ridolfo2012}
\be
\hat O = \sum_{E_n > E_m} \langle \psi_m | (\hat o + \hat o^\dag) | \psi_n \rangle\, | \psi_m \rangle \langle \psi_n|\, ,
\ee
where $|\psi_n \rangle$ ($n = 0\,, 1\,, 2\, \dots$) are the eigenvectors of $\hat H_{\rm s}$ and $E_n$ the corresponding eigenvalues. The superoperator ${\cal D}$ is defined as
\be
{\cal D}[\hat O]\, \hat \rho = \frac{1}{2} (2 \hat O \hat \rho\, \hat O^\dag - \hat O^\dag\hat O\, \hat \rho - \hat \rho\, \hat O^\dag \hat O)\, .
\ee 

The spectrum and the eigenstates of  $\hat H_{\rm s} $ are
obtained by standard numerical diagonalization in a truncated finite-dimensional Hilbert space. The truncation is realized by only including the lowest-energy $N_{\rm c}$ photonic  and $N_{\rm m}$ mechanical Fock states. These truncation
numbers are chosen in order to ensure that the lowest $M < N_{\rm c} \times N_{\rm m}$ energy eigenvalues and the corresponding eigenvectors, which are involved in the dynamical processes investigated here, are not significantly affected when increasing $N_{\rm c}$ and $N_{\rm m}$.
Then the density matrix in the basis of the system eigenstates is truncated in order to exclude all the higher-energy eigenstates which are not populated during the dynamical evolution. This truncation, of course, depends on the excitation strength ${\cal F}(t)$ in Eq.~(\ref{F}).
The system of differential equations resulting from the master equation is then solved by using  a standard Runge-Kutta method with step control. In this way, the mechanical-optical quantum correlations are taken into account to all significant orders.

In writing the master equation, we have assumed that the baths are at zero temperature. The generalization to $T\neq0$ reservoirs can be derived following Ref.~\cite{Hu2015}.
Note that the photonic and mechanical  lowering operators only involve transitions from higher to lower energy states. If $\hat V_{\rm DCE}$ is neglected, $\hat A = \hat a$ and $\hat B = \hat b - (g/\omega_{\rm m}) \hat a^\dag \hat a$. Following Ref.~\cite{Ma2015}, the master equation (\ref{ME}) has been derived without making the post-trace rotating-wave approximation used in Ref.~\cite{Beaudoin2011}, which is not applicable in the presence of equally spaced (even approximately) energy levels, as in the present case.

\subsection{DCE Matrix Elements}
\label{A}

\begin{figure*}
	\centering
	\includegraphics[width = \linewidth]{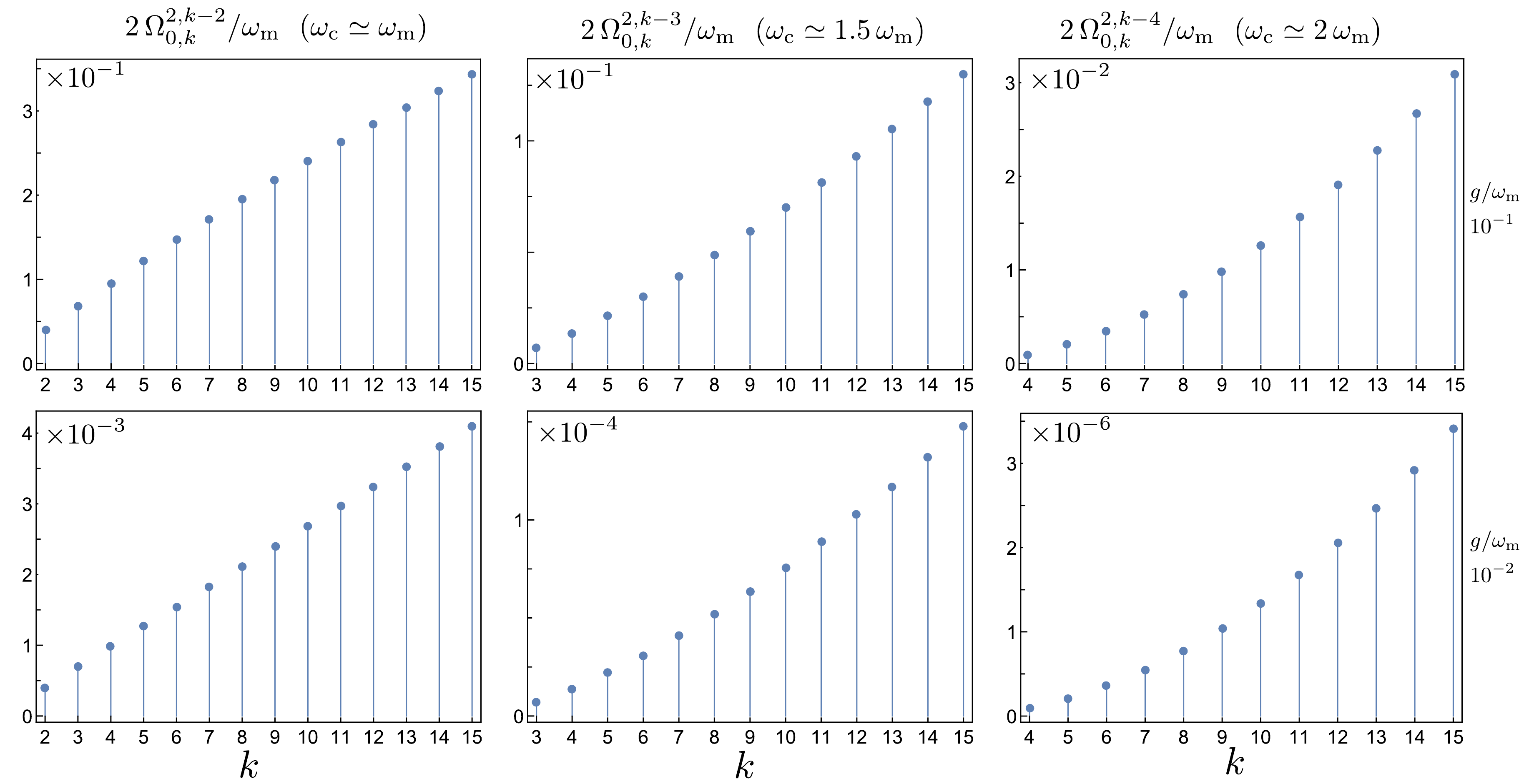}
	\caption{Normalized effective vacuum Rabi splittings $2 \Omega^{2, k-q}_{0,k} / \omega_{\rm m}$ of $\hat V_{DCE}$ between the eigenstates of $\hat H_0 + \hat V_{\rm om}$, evaluated for  $q =2,\, 3,\, 4$, as a function of $k$.
		\label{fig:7}}
\end{figure*}

The matrix elements $\hbar \Omega_{0,k}^{2,k-q} = \langle 0, k| \hat V_{\rm DCE} | 2, (k-q)_2 \rangle$ play a key role in the MDCE, since they determine the rate at which a mechanical Fock state $|k \rangle_{\rm m}$ can generate a photon pair. Figure~\ref{fig:7} displays $2 \hbar \Omega_{0,k}^{2,k-q} = 2 \langle 0, k| \hat V_{\rm DCE} | 2, (k-q)_2 \rangle$ evaluated for  $q =2,\, 3,\, 4$, as a function of the initial Fock state $k$, obtained for  $g/ \omega_{\rm m} = 0.1$ (upper panels) and $0.01$ (lower panels). The two panels on the left, obtained for $q=2$, correspond to the approximate resonance condition $\omega_{\rm c} \simeq \omega_{\rm m}$. The central panels  correspond to $\omega_{\rm c} \simeq  1.5 \omega_{\rm m}$, and the panels on the left correspond to $\omega_{\rm c} \simeq  2 \omega_{\rm m}$.  Going from left to right, the matrix elements decrease. However, as long as they are comparable to the mechanical and photonic decay rates, a mechanical-optical energy exchange  (at least partial) can occur, giving rise to the DCE. 

\begin{figure*}
	\centering
	\includegraphics[width = \linewidth]{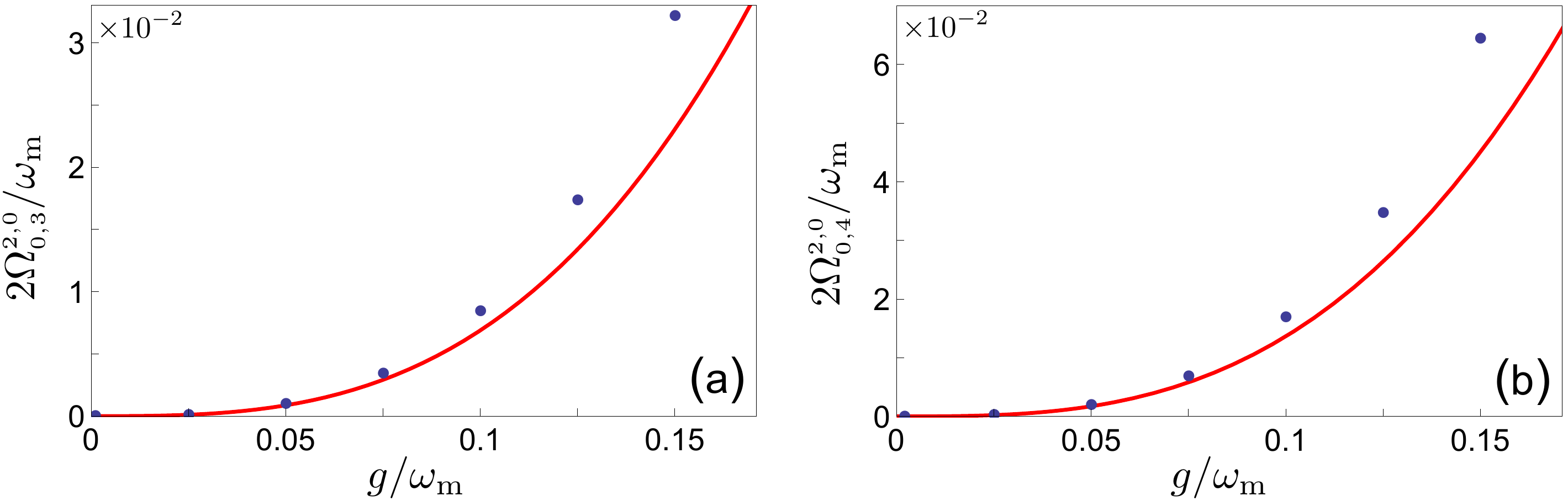}
	\caption{Comparison between the numerically-calculated normalized vacuum Rabi splitting (points) and the corresponding analytical calculations (red solid curve), obtained using first-order perturbation theory  for (a) $2 \Omega_{0,3}^{2,0}$, and (b)  $2 \Omega_{0,4}^{2,0}$. 
		\label{fig:8}}
\end{figure*}

The analytically calculated matrix elements (\ref{V}) displayed in Fig.~\ref{fig:7} describe the phonon-photon DCE coherent couplings obtained using first-order perturbation theory. In order to test their accuracy, we compared them to the corresponding vacuum Casimir-Rabi splittings obtained by numerical diagonalization of $\hat H_{\rm s}$ in Eq.~(\ref{Hs}).
Specifically,  Fig.~\ref{fig:8} shows a comparison for   $2 \Omega_{0,3}^{2,0}$ [panel (a)] and $2 \Omega_{0,4}^{2,0}$ [panel (b)], as a function of the normalized optomechanical coupling $g / \omega_{\rm m}$. The agreement is very good for $g / \omega_{\rm m}$ below 0.1.

\subsection{Negativity}

Negativity is an entanglement monotone and does not increase under local operations and classical  communication \cite{Vidal2002}.
Hence, it represents a proper measure of entanglement, although it can be zero even if the state is entangled, for a specific class of entangled states \cite{Vidal2002}. The negativity of a subsystem A can be defined as  the absolute sum of the negative eigenvalues of the partial transpose $\rho^{T_A}$ of the density matrix $\rho$ with respect to a subsystem A: ${\cal N}(\rho) = \sum_i (|\lambda_i| - \lambda_i)/2$, where $\lambda_i$ are the eigenvalues of $\rho^{T_A}$. In this case, the subsystems A and B are the cavity field and the vibrating mirror, respectively.

	\bibliography{DCE}
	
\end{document}